\documentclass[twocolumn]{aastex63}

\newcommand{\dd}{\textrm{d}}

\shorttitle{}
\shortauthors{.}

\usepackage{tabularx}
\usepackage{amsmath}

\begin{document}
 
\title{Particle In Cell Simulations of Mildly Relativistic Outflows in Kilonova Emissions}

\newcommand{\clf}[1]{\textcolor{blue}{#1}}


\author[0000-0002-8016-400X]{Mohira Rassel}
\affiliation{Center for Theoretical Astrophysics, Los Alamos National Laboratory, Los Alamos, NM, 87545, USA}
\affiliation{Center for Non Linear Studies, Los Alamos National Laboratory, Los Alamos, NM, 87545, USA}

\author[0000-0002-8906-7783]{Patrick Kilian}
\affiliation{Center for Theoretical Astrophysics, Los Alamos National Laboratory, Los Alamos, NM, 87545, USA}
\affiliation{Space Science Institute, 4765 Walnut St, Suite B, Boulder, CO 80301, USA}

\author[0000-0002-2310-4780]{Vito Aberham}
\affiliation{Institut für Theoretische Astrophysik, Universität Heidelberg, 69120 Heidelberg, Germany}

\author[0000-0001-6802-4744]{Felix Spanier}
\affiliation{Institut für Theoretische Astrophysik, Universität Heidelberg, 69120 Heidelberg, Germany}

\author[0000-0003-1707-7998]{Nicole Lloyd-Ronning}
\affiliation{Computational Physics and Methods Group, Los Alamos National Laboratory, Los Alamos, NM 87545, USA}
\affiliation{Department of Math, Engineering \& Science, University of New Mexico,  Los Alamos, NM 87545, USA}

\author[0000-0003-2624-0056]{Chris L. Fryer}
\affiliation{Center for Theoretical Astrophysics, Los Alamos National Laboratory, Los Alamos, NM, 87545, USA}
\affiliation{Computer, Computational, and Statistical Sciences Division, Los Alamos National Laboratory, Los Alamos, NM, 87545, USA}
\affiliation{Center for Non Linear Studies, Los Alamos National Laboratory, Los Alamos, NM, 87545, USA}
\affiliation{The University of Arizona, Tucson, AZ 85721, USA}
\affiliation{Department of Physics and Astronomy, The University of New Mexico, Albuquerque, NM 87131, USA}
\affiliation{The George Washington University, Washington, DC 20052, USA}



\begin{abstract}

The electromagnetic emission from neutron star mergers is comprised of multiple components. Synchrotron emission from the disk-powered jet as well as thermal emission from the merger ejecta (powered by a variety of sources) are among the most studied sources.  The low masses and high velocities of the merger ejecta quickly develop conditions where emission from collisionless shocks becomes critical and synchrotron emission from the merger ejecta constitutes a third component to the observed signal.
The aim of this project is to examine shock development, magnetic field generation and particle acceleration in the case of mildly relativistic shocks, which are expected when the tidal ejecta of neutron star mergers drive a shock into the external medium. Using LANL's VPIC (vector particle-in-cell) code, we have run simulations of such mildly-relativistic, collisionless, weakly-magnetized plasmas and compute the resultant magnetic fields and particle energy spectra. We show the effects of varying plasma conditions, as well as explore the validity of using different proton to electron mass ratios in VPIC. Our results have implications for observing late-time electromagnetic counterparts to gravitational wave detections of neutron star mergers.

\end{abstract}



\section{Introduction} \label{sec:intro}

The merger of compact binaries composed either of two neutron stars or a neutron star and a black hole have, for the past 4 decades, been invoked to explain a wide range of astrophysical phenomena including short-duration gamma-ray bursts ~\citep{1991AcA....41..257P,1999ApJ...518..356P}, sources of r-process yields~\citep{1974ApJ...192L.145L,1979ApJ...228..881N} and origins of gravitational waves~\citep{1963icse.book.....C,1977ApJ...215..311C,1979A&A....72..120C}.  A combination of theory and observations have been used to strengthen these claims.  For example, early predictions for the offset of short-duration gamma-ray bursts under the neutron star merger paradigm~\citep{1999ApJ...526..152F,1999MNRAS.305..763B} ultimately were confirmed by observations, cementing this progenitor for short-duration gamma-ray bursts~\citep{2013ApJ...776...18F}.  Similarly, increasingly detailed studies both of neutron star merger events~\citep{2004NewA....9....1D} and galactic chemical evolution~\citep{2004A&A...416..997A,2016ApJ...830...76K,2017ApJ...836..230C,2019ApJ...875..106C} have shown that these mergers can be the dominant site of r-process production, identifying both the strengths and weaknesses of this r-process source.

The concurrent detection of GW170817 in both gravitational and electromagnetic waves~\citep{2017ApJ...836..230C} has dramatically strengthened this picture. Observing the gravitational waves allowed us to identify a merger event. The combined gamma-ray and radio observations provided strong evidence that this merger produced a relativistic outflow.  The strong infrared signal also suggests that the merger ejected r-process elements, but with a large range of inferred yields~\citep{2017ApJ...836..230C}.  A broad range of physical uncertainties make it difficult to produce quantitative estimates of the r-process ejecta from kilonova observations including the properties of the ejecta and the physics that shapes it and the kilonova emission~\citep[e.g.][]{2018MNRAS.478.3298W,2018ApJ...863L..23Z,2019ApJ...880...22W,2020MNRAS.493.4143F,2020ApJ...899...24E,2021ApJ...918...44B,2021ApJ...906...94Z,2022MNRAS.513.5174P}.  One uncertainty originates in the different sources driving the emission from these mergers, which include radioactive decay, magnetar activity, fallback as well as jet interactions.  

As the gamma-ray burst's jet decelerates, its synchrotron-driven afterglow can contribute to emission over a broad spectral range. However, the jet is not the only source of emission. Thermal emission from the material ejected in the merger (both dynamically and through a disk wind) is believed to be powered by the decay of radioactive isotopes in the outflow.  Additional energy sources for this thermal emission exist.  For example, a contribution by a magnetar/pulsar is possible if the merged neutron-star core develops strong magnetic fields and additionally does not collapse to form a black hole~\citep{2014MNRAS.439.3916M,2019MNRAS.483.1912P,2019ApJ...880...22W}.  Another energy source arises from fallback accretion.  Ejecta moving at below the escape velocity eventually decelerates and falls back onto the compact remnant.  The potential energy released as this material accretes provides an additional energy source~\citep{2019MNRAS.483.1912P,2019ApJ...880...22W}.  Finally, interactions between the jet and wind ejecta convert kinetic energy to thermal energy and this shock heating turns into a substantial contribution to the thermal emission ~\citep{2021MNRAS.500.1772N}. Disentangling these different sources requires combined detailed calculations of broadband electromagnetic emission during the transient higher state. As the ejecta becomes optically thin, the emission from these different sources is not reprocessed and can contribute to the non-thermal emission.

A less-studied source is the non-thermal emission from the deceleration of the ejecta shock~\citep{2008ARA&A..46...89R,2011Natur.478...82N,2015PhRvD..92d4028K,2016ApJ...831..190H,2018ApJ...858...53L}.  Like the gamma-ray burst afterglow, the ejected material will decelerate as it moves through the interstellar medium surrounding the merger and, at the front edge of this decelerating ejected material, a shock is formed that can cause non-thermal emission.  At first glance, this ejecta appears similar to the shock produced in supernova remnants, since it is more isotropic though less energetic than typical gamma-ray burst jets that give rise to afterglow emission. Nevertheless, the velocities of this ejecta tend to be a lot higher than typical supernova remnants.  In this paper, we study the particle acceleration at the barely relativistic shock to non-thermal energies and the resulting synchrotron emission.

 Similar to supernovae, the ejecta shock can produce non-thermal electrons that still power emission long after thermal emission has dissipated. Both the low masses and high velocities of the ejecta from neutron star mergers indicate that this ejecta begins to decelerate, thus entering the kilonova "remnant" phase much sooner than ordinary supernovae.  In this paper, we couple hydrodynamic simulations of the ejecta (Section~\ref{sec:hydro}) with particle-in-cell (PIC) calculations of the particle acceleration within the shock to study the neutron-star merger emission.  Section~\ref{sec:floats} describes our PIC calculations and their results while Section~\ref{sec:synchrotron} deals with the synchrotron emission from these models.  We conclude with a discussion of the importance of these observations and the prospects of future observations.

\section{Kilonova Shock Conditions}
\label{sec:hydro}

To guide our PIC calculations, we must understand the conditions in the shock generated by a kilonova.  Depending on the features of the compact binaries, the ejecta properties (mass, velocity, composition) from a neutron star merger can vary dramatically~\citep[for reviews, see][]{2006ApJ...651..366K,2017Natur.551...80K,2018ApJ...865L..21K,2019ARNPS..69...41S,2021ApJ...918...10W}.   Velocities range from 0.05 to 0.5 times the speed of light (although most lie in the 0.1-0.3c range).  Detailed calculations of these disks~\citep[e.g.][]{2019PhRvD.100b3008M,2021ApJ...906...98N} have yet to fully model the properties of the magneto-radiation-hydrodynamics disks and the full evolution of the disk outflows remains an active area of research.  Typical ejecta masses are typicall between $0.001-0.01\,M_\odot$~\citep{2019ApJ...875..106C}.  Because of both the expected range of and uncertainties in the ejecta properties, our study will span a range of kilonova properties both in mass and peak velocities.

As the shock progresses through the interstellar medium surrounding the binary merger event, the properties of that circumbinary medium also affect our shock properties.  Because of the kicks imparted on neutron stars at birth, the merger can happen far off of the Galactic plane where the density of the surrounding medium is low.  In agreement with many observational estimates of short-duration gamma-ray bursts, galactic evolution calculations of these binaries have demonstrated that these mergers can occur in a wide range of circumbinary densities: ${10^{-4}-10^2\;{\rm cm^{-3}}}$ \citep{wiggins18,2010ApJ...708....9F}.

To study this shock evolution and obtain a first pass at the shock conditions (which we use to produce the initial conditions in our PIC calculations), we use the 1D Lagrangian hydrodynamics code initially designed for supernovae~\citep{fryer99}, modified (using a simple $\gamma_{\rm EOS}=5/3$ equation of state) to follow the ejecta out to large distances.  Although this code assumes collisional equilibrium conditions (Euler assumptions), as we shall see in Section~\ref{sec:results}, turbulent magnetic fields that develop in the shock cause conditions that approximately mimic these Euler conditions. With this code, we calculate a range of kilonova models varying the ejecta mass, peak velocity and circumbinary density.

\begin{figure}[htbp]
\includegraphics[width=3in]{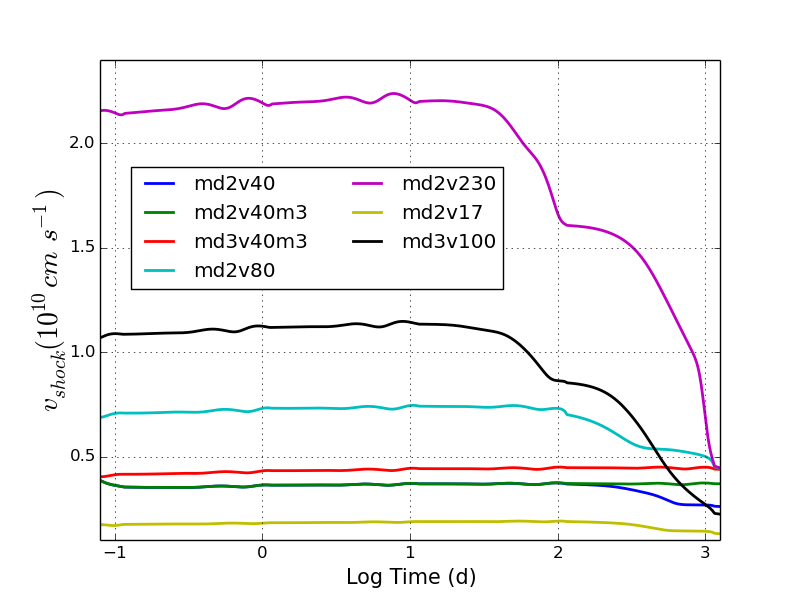}
    \caption{Velocity as a function of time for 7 representative models from our study (Table ~\ref{tab:thermmodels}).  The initial velocity is determined by the binary properties and the conditions in the merger disk.  The shock velocity is ballistic until the ejecta sweeps up sufficient mass to cause deceleration.  In most cases, this occurs beyond 50-100d.}
    \label{fig:vshock}
\end{figure}

Our grid of models is listed in Table~\ref{tab:thermmodels}. This grid spans the ejecta velocities, ejecta masses and circumstellar densities discussed above. The shock velocity as a function of time for a selected set of those models~\ref{fig:vshock}.  Just as with a supernova remnant~\citep{2008ARA&A..46...89R}, the peak ejecta velocity is ballistic until it sweeps up sufficient matter to decelerate the shock.  The time when this occurs depends on both the ejecta mass (increasing with mass), ejecta velocity (decreasing with velocity) and circumbinary density (decreasing with density).  For most of our models, this deceleration occurs after 50d and, prior to this time, the Lorentz factor lies in the range: $1.01 < \Gamma < 1.5$ with more characteristic values of:  $1.1 < \Gamma < 1.3$.

\begin{table}
  \centering
  \scriptsize
  \begin{tabular}{lcccc}
  \hline
  \hline
Model & $m_{\rm ejecta}$ & $v_{\rm shock}$ & $\rho_{\rm CSM}$  \\
 & ($M_\odot$) & ($10^{10} {\rm cm s^{-1}}$) & ${\rm cm^{-3}}$ \\
\hline
 md3v17 & 0.001 & 0.17 & 10 & \\
 md3v34 & 0.001 & 0.34 & 10 & \\
 md3v50 & 0.001 & 0.50 & 10 & \\
 md3v70 & 0.001 & 0.70 & 10 & \\
 md3v80 & 0.001 & 0.80 & 10 & \\
 md3v90 & 0.001 & 0.90 & 10 & \\
 md3v40rm1 & 0.001 & 0.40 & 0.1 \\
 md3v40rm3 & 0.001 & 0.40 & 0.001 \\
 md3v80rm1 & 0.001 & 0.80 & 0.1 \\
 md2v40 & 0.01 & 0.40 & 10 \\
 md2v80 & 0.01 & 0.80 & 10 \\
 md2v120 & 0.01 & 1.20 & 10 \\
 md2v150 & 0.01 & 1.50 & 10 \\
 md2v230 & 0.01 & 2.30 & 10 \\
 md2v40rm1 & 0.01 & 0.40 & 0.1 \\
 md2v40rm3 & 0.01 & 0.40 & 0.001 \\
 
  \end{tabular}
  \caption{Range of the hydrodynamic models that we use to determine realistic conditions at the shock front for the particle-in-cell simulations. We vary the total mass of the ejecta, the initial shock velocity and the number density of the unshocked circumbinary material.}
  \label{tab:thermmodels}
\end{table}

\begin{figure}[htbp]
\includegraphics[width=3in]{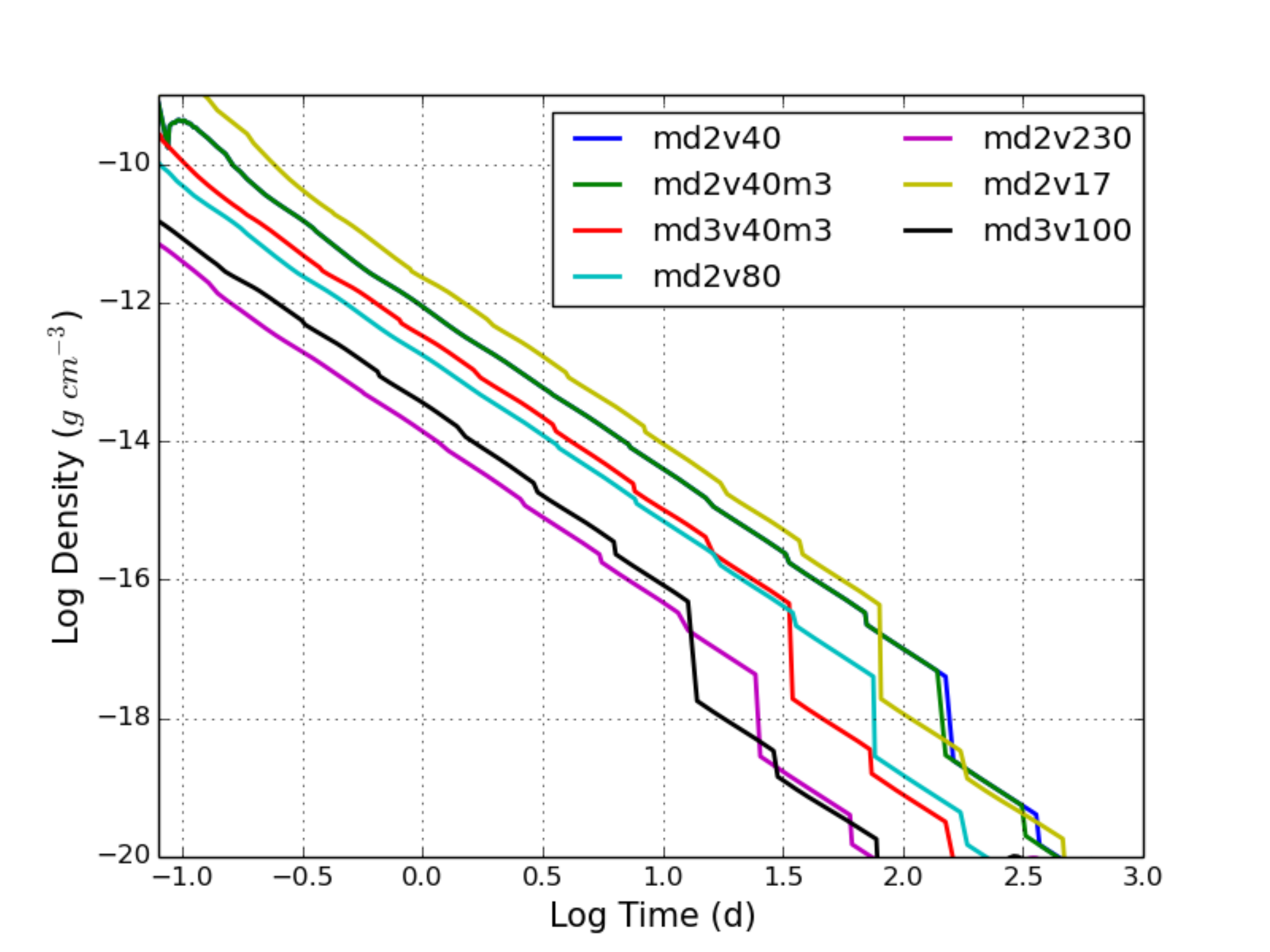}
    \caption{Density at the shock front as a function of time for the suite of models shown in Figure~\ref{fig:vshock}.  The density at this shock front decreases roughly with time:  $\rho_{\rm shock} \propto t^{-2}$.}
    \label{fig:rhoshock}
\end{figure}

The corresponding density at the shock front as a function of time is affected by this deceleration, but is primarily set by the density of the circumbinary medium.  In our suite of models listed in Table~\ref{tab:thermmodels}, we varied the circumbinary density from 0.001-10 particles cm$^{-3}$.  The density at the shock front as a function of time is shown in Figure~\ref{fig:rhoshock}.  The shock density decreases with time:  $\rho_{\rm shock} \propto t^{-2}$.

\begin{figure}[htbp]
\includegraphics[width=3in]{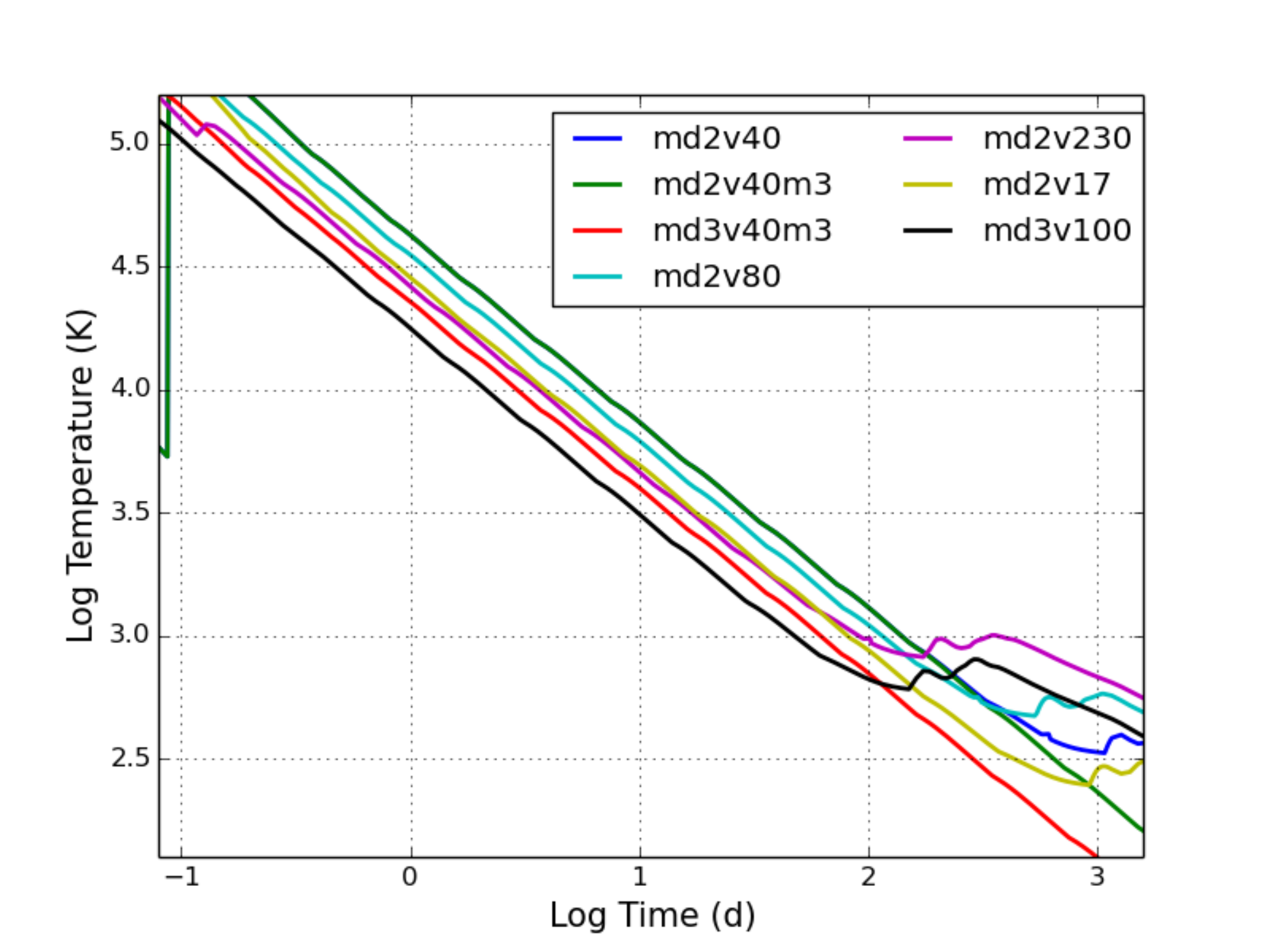}
    \caption{Temperature at the shock front as a function of time assuming the bulk of the energy is converted into radiation energy for the suite of models shown in Figure~\ref{fig:vshock}.  The coupling between electrons and radiation is not perfect and this temperature marks a lower limit on the shock temperature.}
    \label{fig:tempshock}
\end{figure}

Determining the temperature of this shock is more difficult.  To determine the pressure, we assumed the electrons and ions are in equilibrium, described by a Maxwellian.  As we shall see, our PIC calculations show that the magnetic fields generated in the shock can mimic a collisional shock, helping to drive to such an equilibrium.  If we further assume that the electrons and photons equilibrate and the photons are additionally described as a Planckian, we can estimate the temperature of this gas.  Figure~\ref{fig:tempshock} shows the shock temperature as a function of time for our shock models based on these equilibrium assumptions.

\begin{figure}[htbp]
\includegraphics[width=3in]{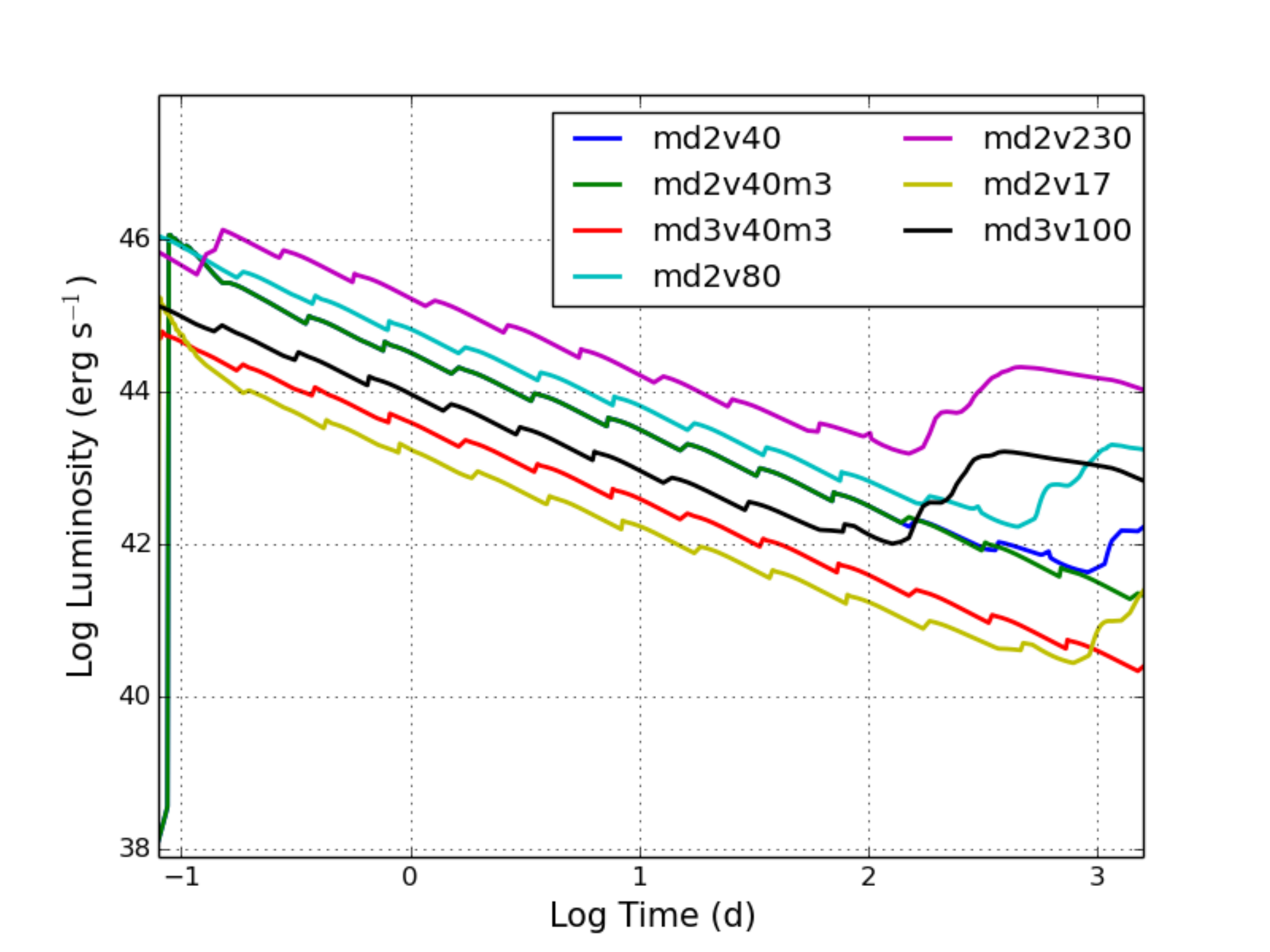}
    \caption{Luminosity of the shock as a function of time assuming the bulk of the energy is converted into radiation energy for the suite of models shown in Figure~\ref{fig:vshock}.}
    \label{fig:eqluminosity}
\end{figure}

The corresponding luminosity from this shock assuming blackbody emission is shown in Figure~\ref{fig:eqluminosity}.  Again, this calculation assumes full coupling between ions, electrons and photons.  In this local thermodynamic equilibrium scenario with the temperatures of all particles equal, the energy from the shock is deposited almost entirely in the radiation.  At such low densities the ions, electrons and photons may not be fully coupled and their energy distributions may not be described by a Maxwellian/Planckian.  Both our spectra and luminosities will therefore be affected by deviations from equilibrium.  

To understand these deviations, we must understand how the shock energy is distributed into different particles.  The shock energy initially flows into the ions that then couple to the electrons, which themselves couple to the radiation. Since the energy distribution of the electrons will be at lower energies than the ions' respective distribution, the ions and electrons are actually not fully coupled.  In addition, the electrons are also not fully coupled to the radiation so that much of the energy remains in the ions and electrons.  Accordingly, the electron's effective temperature is higher and the radiation spectra, which will not follow a blackbody distribution, will likely have much higher energies than those predicted here.  

The goal of this paper is to study these conditions using particle-in-cell calculations to determine the extent at which these equilibrium solutions are appropriate.  We know that a subset of the electrons will be accelerated beyond the energies predicted by these equilibrium and we will discuss the nonthermal emission from these models in Section~\ref{sec:floats}.

\section{Particle-in-Cell Simulations} \label{sec:floats}
We use LANL's VPIC code \citep{Bowers2008} for the particle-in-cell (PiC) simulations. Like most other electromagnetic PiC codes, this code loads a large number of computational particles into a simulation domain that contains a grid for currents and field quantities. The currents from moving particles are calculated and interpolated onto the grid. From those grid conditions, the electric and the magnetic fields are updated using Maxwell's equations. They are then subsequently interpolated to the positions of the particles, which are pushed by the Lorentz force using the Boris push.

\subsection{Simulation Setup}

For this project we do a grid of calculations varying proton-to-electron mass ratios, orientation of the magnetic fields, strength of the magnetic field relative to the energy density (plasma beta: $\beta=$ energy density divided by the magnetic field energy density) and Lorentz factor of the ejecta ($\Gamma$). The fiducial values for these parameters is as follows: The proton-to-electron mass ratios is typically $m_i / m_e = 100$, which is sufficient to get good scale separation between electron and proton physics, including an electrostatic cross-shock potential and little variation of final results to further increased of the mass ratio. The shock-normal angle between the upstream flow direction and the upstream magnetic field is 60 degree, producing oblique shocks with efficient particle acceleration in the barely relativistic regime. 
This is consistent with the assumption that the shock propagates into the interstellar medium with a randomly oriented magnetic field, which is likely to be oblique to the radial direction.
The plasma beta $\beta$ is typically $\beta = 10$, much larger than in most other kinetic simulation studies. This leads to shock velocities on the order of $1.1\cdot10^{10}$cm/s, well within in the range given by Fig.~\ref{fig:vshock}. In the interstellar medium, through which most kilonova explosions propagate, the value for $\beta \approx 1-10$\citep{2008RPPh...71d6901K}.  For gamma-ray burst studies, the post-shock values lie somewhere between 1-100\citep{2002ApJ...571..779P,2005MNRAS.363.1409P}.  The flow speed of the the upstream is typically such that $\Gamma = 1.047$, lower than most previous studies. This is realistic for later stage outflow, but comes at an increased computational cost for the shock to form and the particle distribution to reach steady state.
Consequently our simulation set-up is quasi 1D. Only the $x$ direction of $L_x = 1800 d_i$ is resolved by $N_x = 38912$ cells. The $z$ direction is $L_z = 0.049 d_i$ and is resolved by just $N_y = 8$ cells. The $y$ direction is unresolved and described by a single cell. The cell size is sufficient to resolve the Debye length and all larger kinetic length scales. In velocity space, all three components of the particle velocities retain all three components as well as the electromagnetic fields and currents. As such, we term these simulations 1D3V.  The left boundary at $x=0$ is set to be a reflecting, conducting wall in the downstream reference frame. Differing from its left counterpart, the right boundary is set as an absorbing wall while the walls in the $y$ and $z$ directions are set periodic, allowing particles to reappear on the other side of the boundary if they leave the simulation's boundaries.
Particles are loaded with a Maxwellian distribution of temperature $k_B T = 0.005\,m_{e}c^{2} = 2.555\mathrm{keV}$. This is somewhat higher than the temperatures commonly assumed for the hot ionized medium (0.1 -- 1keV) but increases the cell size and time step length to reduce the computational cost without affecting the physics results unduly. All particles initially belong to the upstream and are given mildly relativistic upstream bulk flow speed
in the negative x-direction towards the reflecting wall. For each case, 100 particles per cell per species are used.
The time-step $\Delta{t}$ is set to $0.9$ of the maximum allowed by the CFL condition in two dimensions.
All runs are performed in normalized units, which includes setting the initial upstream density to unity in VPIC. This is permissible since our plasma is collisionless and no processes that are non-linear in the plasma density are modeled here. To convert results to physical units in post-processing we need to pick a realistic value for this density from Fig.~\ref{fig:rhoshock}.

We vary the input parameters around our fiducial run.  As with many PIC calculations of shocks, we lower the proton to electron mass ratio to reduce simulation cost, but vary this ratio using 3 values (42, 100, 200) to test its importance.  We have also considered a range of different oblique angles between the magnetic fields and the $x$ axis, using $45^\circ$, $55^\circ$, $60^\circ$, $72^\circ$ and $90^\circ$. We did runs with different magnetic field, resulting in $\beta$ =  10, 30, 50 and we modified the upstream flow speed corresponding to $\Gamma$=1.047, 1.1 and 1.3 in different runs. The Numerical Cherenkov Instability (NCI) is not of major concern here since our velocities are only mildly relativistic and the fraction of highly-relativistic, non-thermal particles remains small.

\subsection{Electron Distribution} 
\label{sec:results}

\begin{figure}[htbp]
\includegraphics[width=2.75in]{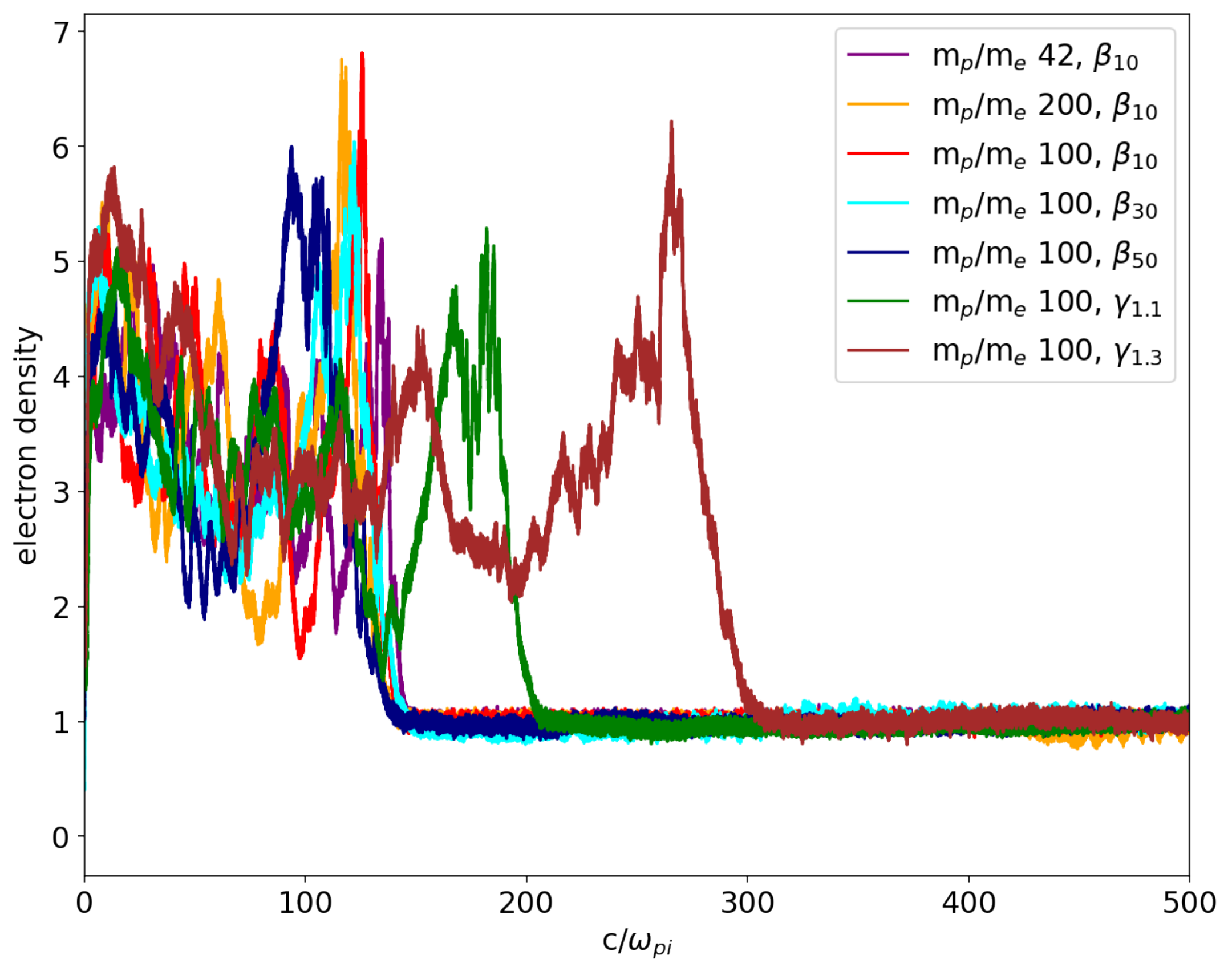}
    \caption{Electron density profile for all mass ratios, $\beta$ and $\Gamma$ for the same timestep. $\beta_{0.5}$ has the most relaxed downstream due to a stronger initial magnetic field. $\Gamma_{1.1}$ and $\Gamma_{1.3}$ have the largest downstream due to initial upstream velocities.  }
    \label{fig:elecDen}
\end{figure}

\begin{figure}[htbp]
\includegraphics[width=3in]{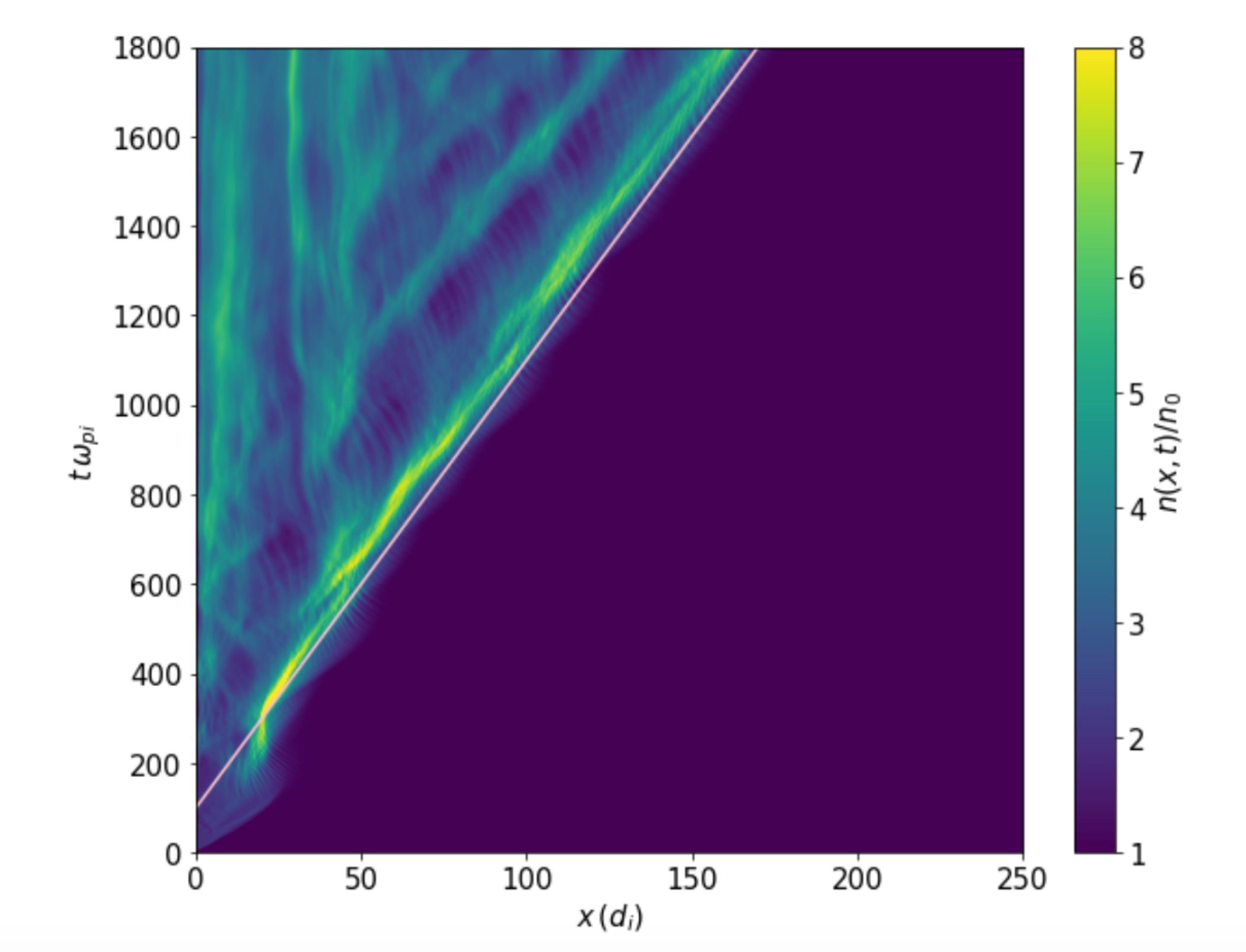}
    \caption{$n(x,t)$ plot for $m_{p}/m_{e}$=100 at $60^\circ$. The x axis is in the ion inertial length. y axis is time $\times$ ion plasma frequency. The shock forms at around $300\omega_{pi}^{-1}$. After that the shock front moves a nearly constant speed and the downstream shows the typically ripples.}
    \label{fig:nxt}
\end{figure}

One of the key results of our calculations is the resultant nonthermal electron energy distribution, usually fit by a power-law.  
Spectral indices of p=-4 are expected for non-relativistic diffusive shock acceleration (DSA) where we expect $p = -3r/(r-1)$ \citep{Bell_1978,Blandford_1978, Drury_1983, Blandford_1987}.
For increasingly relativistic shocks with larger $\Gamma$ (see below) we expect this to harden since Fermi acceleration becomes more important that leads to $p=-(r+2)/(r-1)\approx 2$.  In this section, we review our simulation grid, focusing on the power-law index of the electron energy distribution. 

The results from our grid of models are summarized in Table~\ref{tab:42_10} and  Table~\ref{tab:60}.
Results with varying oblique angles are shown in Table~\ref{tab:42_10} and results for oblique angle $60^\circ$ are shown in Table~\ref{tab:60}.
The $60^\circ$  angle produces the most-pronounced non-thermal tail of electrons with a relatively hard spectrum and is close to the critical angle that separates subluminal and superluminal shocks. \cite{Sironi_2011} also found a particularly pronounced tail of energetic electrons for $\theta \approx \theta_{crot}$, but at a relativistic upstream $\Gamma$. \cite{Xu_2020} also found that the optimal angle is between 60 and 70 degree, for a non-relativistic shock similar to ours.

All of our simulations produced a well developed shock, which can be seen in Figures \ref{fig:elecDen} and \ref{fig:nxt}. These runs are representative of the results since the values for both $\Gamma$ and $\beta$ are reasonable. As displayed in the obtained electron density profiles $n(x,t)$ and quite similar compression ratios, these results are reasonable.  Figure \ref{fig:elecDen} shows the electron density profiles at timestep 400000.  At this time for all models, a well developed downstream, an overshoot and an unperturbed upstream exists. The shock produces a compression ratio for all simulations that are close to $n_d/n_u = 4$ where $n_d$ is the downstream density and $n_u$ is the upstream density. Such a density jump indicates that we are approaching the strong shock limit, which for an ideal gas with $\gamma_\textrm{EOS}=5/3$ converges to a compression ratio of  $(\gamma_{\textrm{EOS}}+1)/(\gamma_{\textrm{EOS}}-1) = 4$ \citep{balogh2013}. For bulk properties of the downstream, this justifies the hydrodynamic calculations in Section~\ref{sec:hydro}.  But note that the shock profile shows distinct features such as an overdensity right at the shock front itself as well as lower density regions that settle to a steady state over a finite width of the downstream. Emission mechanics and other processes that depend on local density squared (such as e.g. charge exchange collisions) would be sensitive to these variations in density which could lead to a emphasis of the shock front in observed emission.
The shock in the run with $\Gamma=1.1$ has progressed farther since the larger upstream velocity also leads to a larger shock velocity in the downstream frame. This trend continues with that run that as an upstream with $\Gamma = 1.3$.

Figure \ref{fig:nxt} is a stacked plot of density profiles $n(x,t)$ for the run with $m_{p}/m_{e}=100$,  $\beta_{10}$ and $\Gamma = 1.047$. In this plot we can see fluctuations and ripples in the downstream and a constant shock speed of $v_{sh} = 0.109 c$. This shows that this run was run long enough so that it has a well-developed downstream.

To limit the number of simulations, we do not cover the full parameter phase-space.  Instead, we conduct a series of studies where all but one parameter is held fixed.  For example, using a proton to electron mass ratio set to 42 ($\approx\sqrt{1836}$), a plasma beta $\beta = 10$ and $\Gamma = 1.047$, we varied the normal shock angle to examine how the angle of the initial magnetic field affects our powerlaw tails. These values are listed in Table \ref{tab:42_10}.  For these runs, the simulated power-law values are much steeper than used in gamma-ray burst studies (which assume $p\approx-2$), ranging from -3.7 to -11.9.  The flattest power-law value is produced when the magnetic field angle with respect to the shock is $60^\circ$. 

The values of the ratio eB/kE are also higher than the the typical 0.01-0.1 ratio commonly used in the literature.  Since this quasi-normal angle produced the most optimistic results for synchrotron emission, we fixed the angle accordingly (at $60^\circ$) and focused our study on the other parameters.  Table \ref{tab:60} shows the results where we vary both the proton to electron mass ratio and the plasma beta. The results from our mildly relativistic shocks lie in between the non-relativistic and relativistic values.  In addition, increasing the mass ratio directly affects the power law's slope, which correspondingly starts decreasing. There is a drastic change from $m_{p}/m_{e}$=42 to $m_{p}/m_{e}$= 100. The highest mass ratio of all used, $m_{p}/m_{e}$=200 features the hardest powerlaw tail.  

\begin{figure}[htbp]
\includegraphics[width=3in]{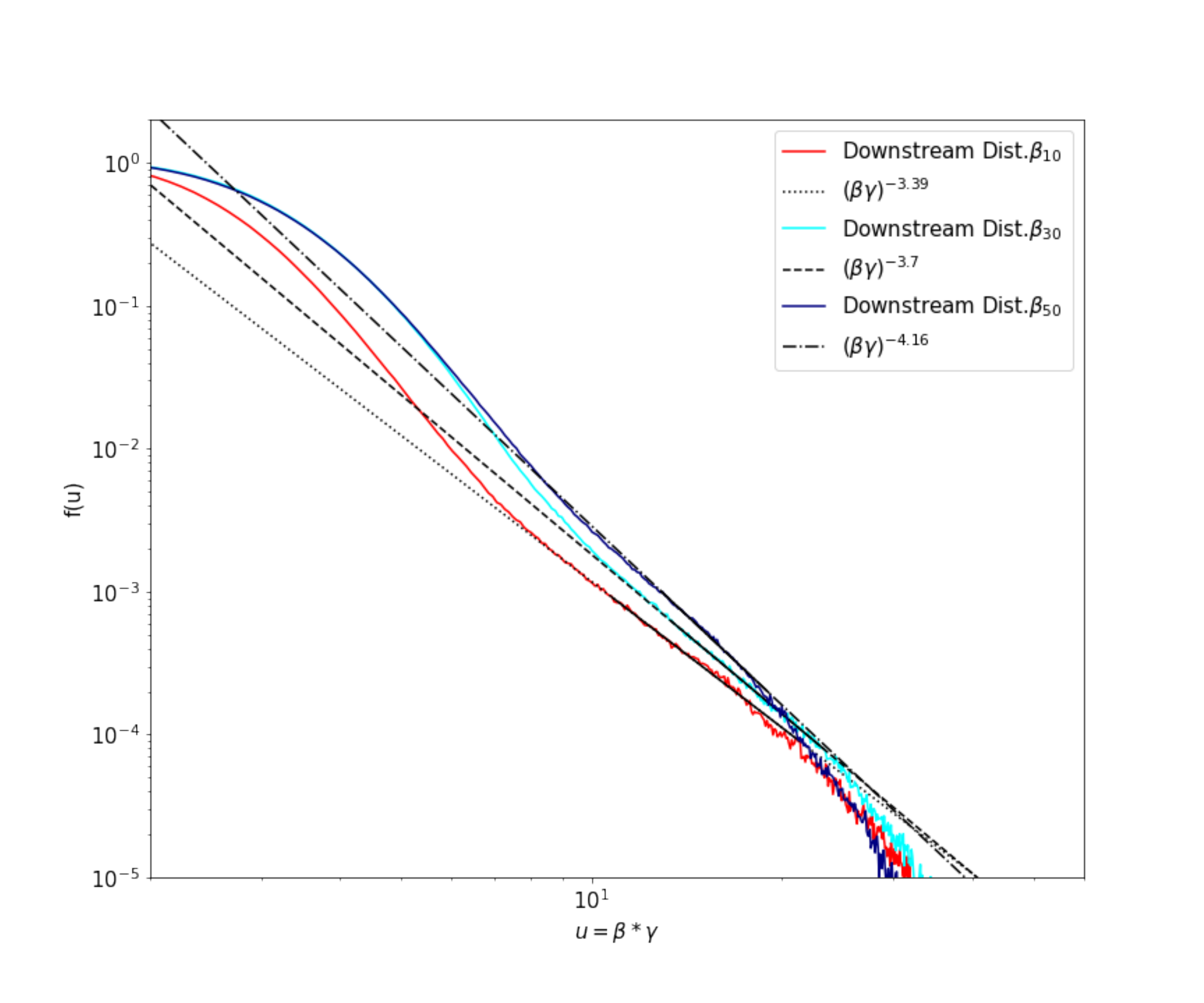}
    \caption{Powerlaw tail fit for $\beta_{10}$, $\beta_{30}$ and $\beta_{50}$ with $m_i/m_e = 100$.  }
    \label{fig:betas}
\end{figure}

For our next group of simulations we chose a fixed mass ratio of 100, varying the values for the plasma beta and the shock Lorentz factor. We examined runs with $\beta$ = 7, 10, 30 and 50. Although we ran simulations for $\beta$ smaller than 7, those aren't as applicable to the environment of our simulations. Figure \ref{fig:betas} shows power-law fit plots for different $\beta$ with the fixed mass ratio of exactly 100. Increasing $\beta$, thus decreasing initial magnetic field, led to steeper slopes. The only exception, the run incorporating $\beta_{3}$ has a larger slope than $\beta_{30}$ and almost the same as $\beta_{50}$. The trend is that if we keep decreasing $\beta$, the obtained powerlaw tail will become steeper. Running simulations in full 3d and for a longer time might modify this trend since current driven instabilities might produce a stronger magnetic field -- capable of scattering particles -- at the shock front in that case.

\begin{figure}[htbp]
\includegraphics[width=2.75in]{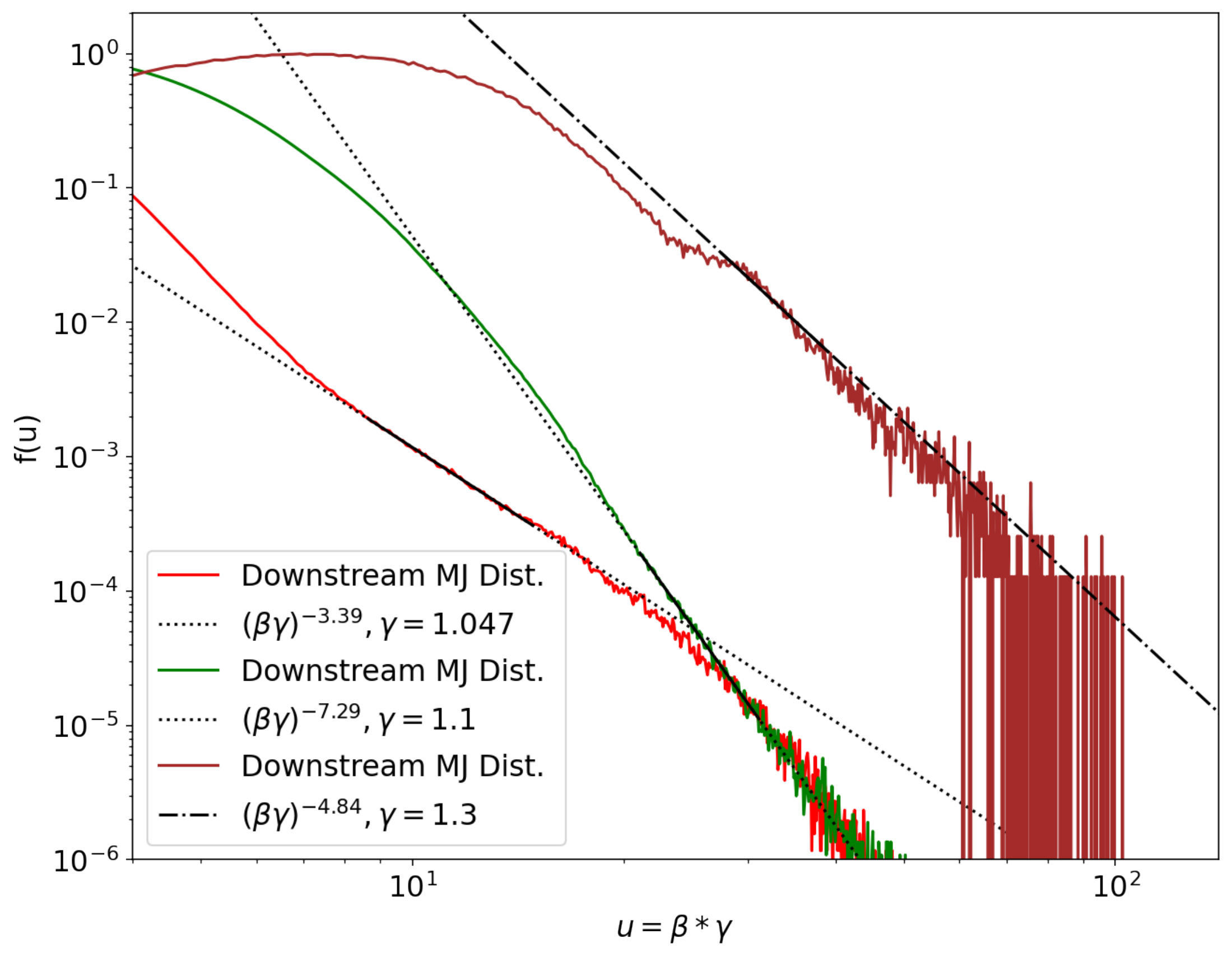}
    \caption{Powerlaw  tail fit for $\Gamma$=1.047, 1.1 and 1.3. Increasing $\Gamma$ caused the overall powerlaw tail to become steeper. -p for  $\Gamma$=1.3 is steeper than $\Gamma$=1.1 due to unexpected intermediate spectral breaks.}
    \label{fig:powerlaw_gammas}
\end{figure}

In the next subsection we will also explore how changing the Lorentz factor $\Gamma$ affects the number of non-thermal particles. For now we focus on the spectral shape.
When comparing $\Gamma$= 1.047, 1.1 and 1.3 the one that has the steepest powerlaw tail is 1.047. We are still in the mildly relativistic regime and larger bulk speeds of the upstream provide additional free energy that can harden the power law distributions of the non-thermal electrons, so we expect that the slops of the runs with $\Gamma$ 1.1 and 1.3 should be shallower than for the run with $\Gamma = 1.07$. This comparison is complicated by the fact that two of the runs have a shallower shoulder at intermediate energies as seen in Figure \ref{fig:powerlaw_gammas}. For the run with $\Gamma = 1.047$ this should has a slope of $p=-3.39$ which softens to $p=-7.3$ in the tail of the distribution. For a higher $\Gamma$ of 1.1 this shoulder vanishes and we get a single powerlaw with $p=-7.29$. Overall this contains more energetic particles without making the tail end of the distribution significantly harder. For $\Gamma = 1.3$ the shoulder appears again, with a $p=-2.1$ (fit not shown) and a tail that has hardened to $p=-4.84$. Additionally the number of energetic particles has increased. Since we find that the distribution functions are not simple power laws, we compute the synchrotron radiation from the full distribution functions instead of making power law approximations. 

\begin{figure}[htbp]
\includegraphics[width=2.5in]{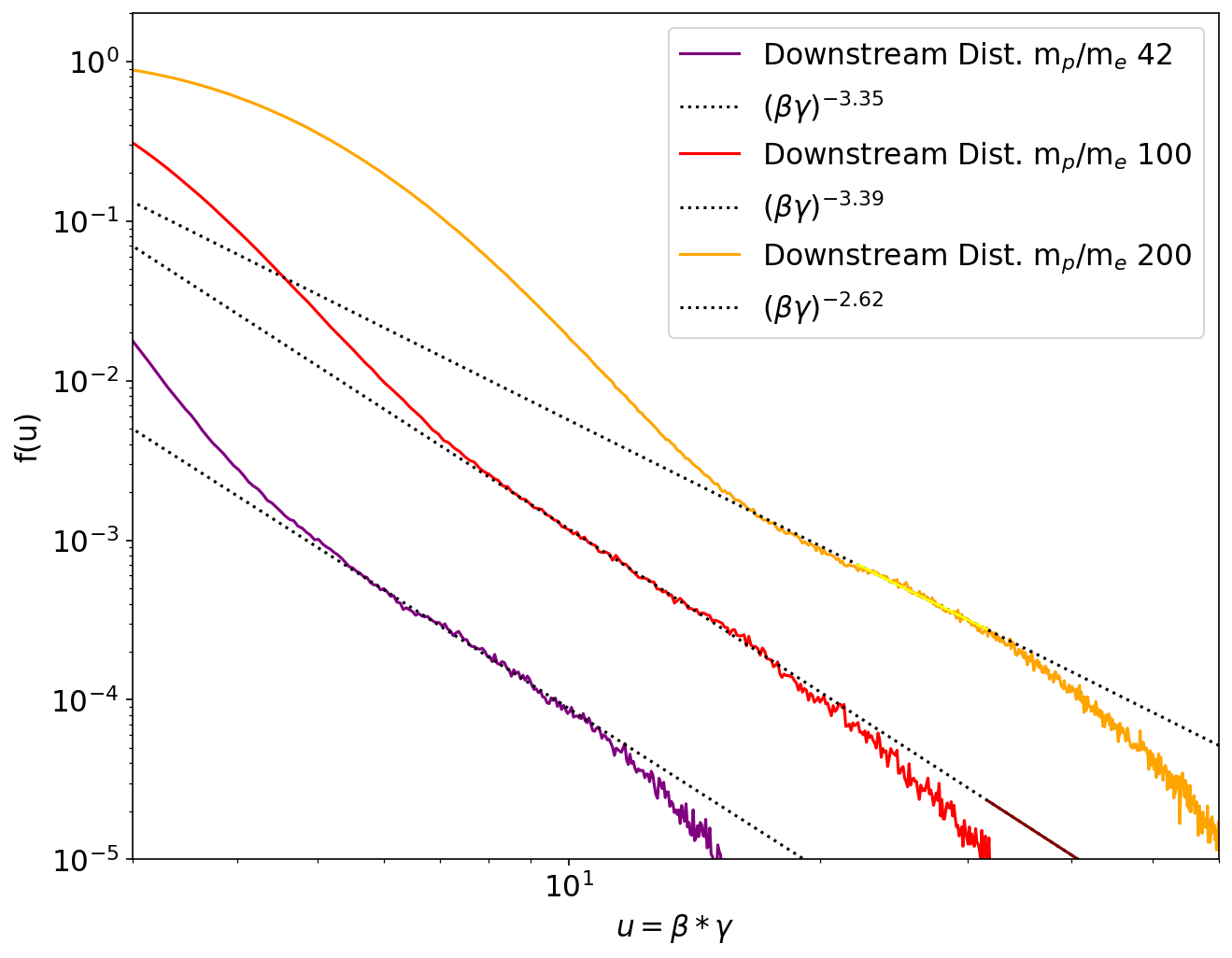}
    \caption{Powerlaw fit for $60^\circ$, $\beta$ 10 with different mass ratios. Apparently the mass ratio matters because increasing it causes a steeper power law tail.}
    \label{fig:powerlaw10}
\end{figure}

\begin{table}
  \centering
  \scriptsize
  \begin{tabular}{lccccc }
  \hline
  normal  &  $B_d/B_u$ & $n_d/n_u$ & $E_B/E_k$ &-p \\
  shock$^\circ$  \\
\hline
 45  & 2.90 & 7.92 & 0.2983 & -7 \\
 55 & 2.75 & 3.14 & 0.2583 & -7 \\

 60 & 3.01 & 3.25  & 0.2561 & -3.73 \\

 72 & 2.51 & 2.31 & 0.2842 & -11.45 \\

 90  & 2.85 & 2.87 & 0.1898 & -11.88 \\
  \end{tabular}
  \caption{Results for runs with $m_{p}/m_{e}$=42. $B_d/B_u$ is the magnetic compression ratio, $n_d/n_u$ is electron density compression ratio. All angles besides $60^\circ$ had steep slopes. $72^\circ$ and $90^\circ$ had the steepest slopes. }
  \label{tab:42_10}
\end{table}

\begin{table}[htbp]
  \scriptsize
  \begin{tabular}{lcccccc}
  \hline
  $m_{p}/m_{e}$ & $\beta$ & $B_d/B_u$ & $n_d/n_u$ & $E_B/E_k$ & -p \\
    \\
\hline
 42  & 10 & 3.31 & 3.08  & 0.235 & -3.63 \\
 
 100 & 7 & 2.89 & 2.96 & 0.383 & -3.34\\
 
 100  & 10 & 3.36 &  3.37 & 3.32 & -3.39 \\

 100  & 30  & 2.27 & 3.702 & 0.1791 & -3.7 \\

 100  & 50 & 1.79 & 3.156  & 0.146 & -4.16 \\

 200  & 10 & 2.96 & 3.25  & 0.3843 & -2.66\\

  \end{tabular}
  \caption{Results for $60^\circ$ angle. $B_d/B_u$ is the magnetic compression ratio while $n_d/n_u$ describes the compression ratio of the electron density. eB/kE is the energy of the average downstream magnetic field divided by the kinetic energy. $B_d^2/8/\pi$ /upstream kinetic energy p particle spectral index.}
  \label{tab:60}
\end{table}

\subsection{Non-thermal Particles}

To calculate the fraction of electrons in the power law, we fit a Maxwell-Jüttner (MJ) Distribution to the low-energy part of our particle spectra to get 
\begin{align}
  f_\textrm{MJ}(u) = \frac{u\,\sqrt{1+u^2}}{\Theta  K_{2}(\frac{1}{\Theta})} e^{-\frac{\sqrt{1+u^2}}{\Theta}}
\end{align}
where $u = \beta \gamma$, $\beta = \sqrt{1-1/\gamma^2}$, $\gamma = 1/\sqrt{1-v^2/c^2}$, relativistic temperature $\Theta = k_B T/mc^2$, $K_{2}$ is the modified Bessel function of the second kind. The Maxwell-Jüttner distribution is then subtracted from the actual distribution function to obtain the distribution function of the high energy (HE) particles: 
\begin{align}
 f_\textrm{HE}(u)=f(u)-f_\textrm{MJ}(u)\,.
\end{align}
All negative values, where the MJ distribution predicts larger numbers than we actually have, are set to zero. After that the MJ distribution is integrated to get the number of particles following the MJ distribution, 
\begin{align}
 N_\textrm{MJ} = \int f_\textrm{MJ}(u)\dd u\,,
\end{align}
before the number of high energy particles is computed analogously: 
\begin{align}
 N_\textrm{HE} = \int f_\textrm{HE}(u)\dd u\,.
\end{align}
The ratio of these two, $N_\textrm{HE}/N_\textrm{MJ}$, is taken ultimately. 

\begin{table}
  \centering
  \scriptsize
  \begin{tabular}{lcccccc}
  \hline
  $m_{p}/m_{e}$ & $\beta$ & $\Gamma$  & $N_{HE}/N_{MJ}$  & $\gamma_{M}$  & $k_B T_{e}/ m_e c^2$  & $k_B T_{i} / m_e c^2$  \\

\hline
 42  & 10 &  1.047  & 0.714  & 4.5 & 0.37 & 0.03 \\

 100  & 10 & 1.047 & 0.230 & 7 & 0.53 &  2.6\\

 100  & 10 &  1.1  &  0.028  & 13 & 1.2 & 6.5\\

 100  & 10 & 1.3 & 0.062 & 42 & 3.8 & 18.0\\
 
 100  & 30 & 1.047 & 0.180 & 7 & 0.7 & 2.6 \\
 
 100  & 50 & 1.047 & 0.317& 7.7 & 0.8  & 3.3\\
 
 200  & 10 & 1.047 & 0.109 & 10 & 1.03 & 6.0 \\

  \end{tabular}
  \caption{Thermal ratio for each run. $k_B T_{e}/ m_e c^2$ is electron temperature in the downstream, $k_B T_{i} / m_e c^2$ is ion temperature in the downstream. }
  \label{tab:thermal}
\end{table}

This ratio changes with changing the Lorentz factor $\Gamma$.
Our hydrodynamic models pointed at Lorentz factors between $\Gamma$=1.01 and $\Gamma$=1.3. We ran these simulations with mass ratio 100. Results for these runs can be seen in Tables~\ref{tab:60} and \ref{tab:thermal}.
In simulations of relativistic shocks with $\Gamma = 15$ \cite{Sironi_2011} found that non-thermal electrons are about 13 percent by number and contain about 35 percent of the electron kinetic energy. In mildly relativistic simulations with $\Gamma = 1.5$ \cite{Crumley_2019} found that the number fraction of non-thermal electrons is about $5\cdot 10^{-4}$. In non-relativistic simulations with $\Gamma = 1.005$ \cite{Park_2015} found a similar number fraction of injected electrons. These numbers however only count particles above a minimum injection energy where we count all excess particles that exist in addition to the number of particles predicted from a Maxwell-Jüttner-distribution fitted to the lower energy part of the distribution. We see a slight drop in $N_{HE}/N_{MJ}$ with increasing $m_p/m_e$, but the ratio is also affected significantly by the choice of plasma $\beta$ so we limit our conclusion to the observation that there can be about 10 percent of particles that have more energy than expected from the thermal distribution that is employed in hydrodynamic simulations.

\section{Synchrotron Radiation} \label{sec:synchrotron}

To determine the spectra and luminosity of our shocks, we calculate the synchrotron emission from the particle distribution generated within the PiC simulation. This is done as a postprocessing step, which in turn means that we do not consider synchrotron cooling within the simulation. This approach is valid since the cooling time scale exceeds the typical simulation time.

The synchrotron emission coefficient is calculated using the Melrose-approximation:
\begin{align}
	j_\nu &= \frac{1}{4\pi} \int_1^\infty \dd \gamma\ n(\gamma) P_{\nu,s}(\gamma)\\
	&\approx \frac{\sqrt{3} B e^3 \beta^2}{4\pi m c^2}\left(\frac{\nu}{\nu_0}\right)^{1/3}\int_0^\infty \dd p\  n(p) \left(1+\left(\frac{p}{mc}\right)^2 \right)^{-1/3}\nonumber\\ &\ \times \exp\left[-\frac{\nu}{\nu_0\left(1+\left(\frac{p}{mc}\right)^2 \right)} \right]\,.\label{eq:melrose}
\end{align}
Electron momentum histograms from the simulation are used to numerically integrate Eq. \eqref{eq:melrose} which yields the synchrotron emissivity right away.
The process of synchrotron emission is accompanied by the possible absorption of the very synchrotron photons by particles in the plasma. This is called synchrotron self-absorption (SSA). The general idea of this process is related to the Einstein coefficients for emission, spontaneous and stimulated emission in a continuum. Further details are given in \citet{rybicki}. The derivation given there is strictly limited to ultra-high energies, which is a scenario not applicable here. A detailed derivation of the SSA coefficient $\alpha_{\nu,\textrm{SSA}}$ and corresponding optical thickness $\tau_{\nu,\textrm{SSA}}$ suitable for the mildy relativistic setting is given in Appendix \ref{app:alpha}.

Synchrotron emissivity ($j_\nu$) and optical thickness are combined assuming emission from a shell with radius $R_\text{S}$, a thickness $d_\text{shock}$, and a  luminosity distance $l_\text{D}$. To get a realistic value for $R_\text{S}$ we use the results from the hydrodynamics code which indicate that at one day after the kilonova $R_\text{S} \approx 2\cdot10^{15}$ -- $2\cdot10^{16}\text{cm}$ which expand to about $3\cdot10^{17}\text{cm}$ by the third day.
The spectral energy distribution is then accordingly given as
\begin{align}
    F_\nu &= \frac{4\pi R_\text{S}^2}{l_\text{D}^2}\times d_\text{shock} \times j_\nu \times \exp\left[ -\alpha_{\nu,\textrm{SSA}} \times d_\text{shock} \right]\,.
    \label{eq:Fnu}
\end{align}

\begin{figure}[htbp]
\includegraphics[width=3in]{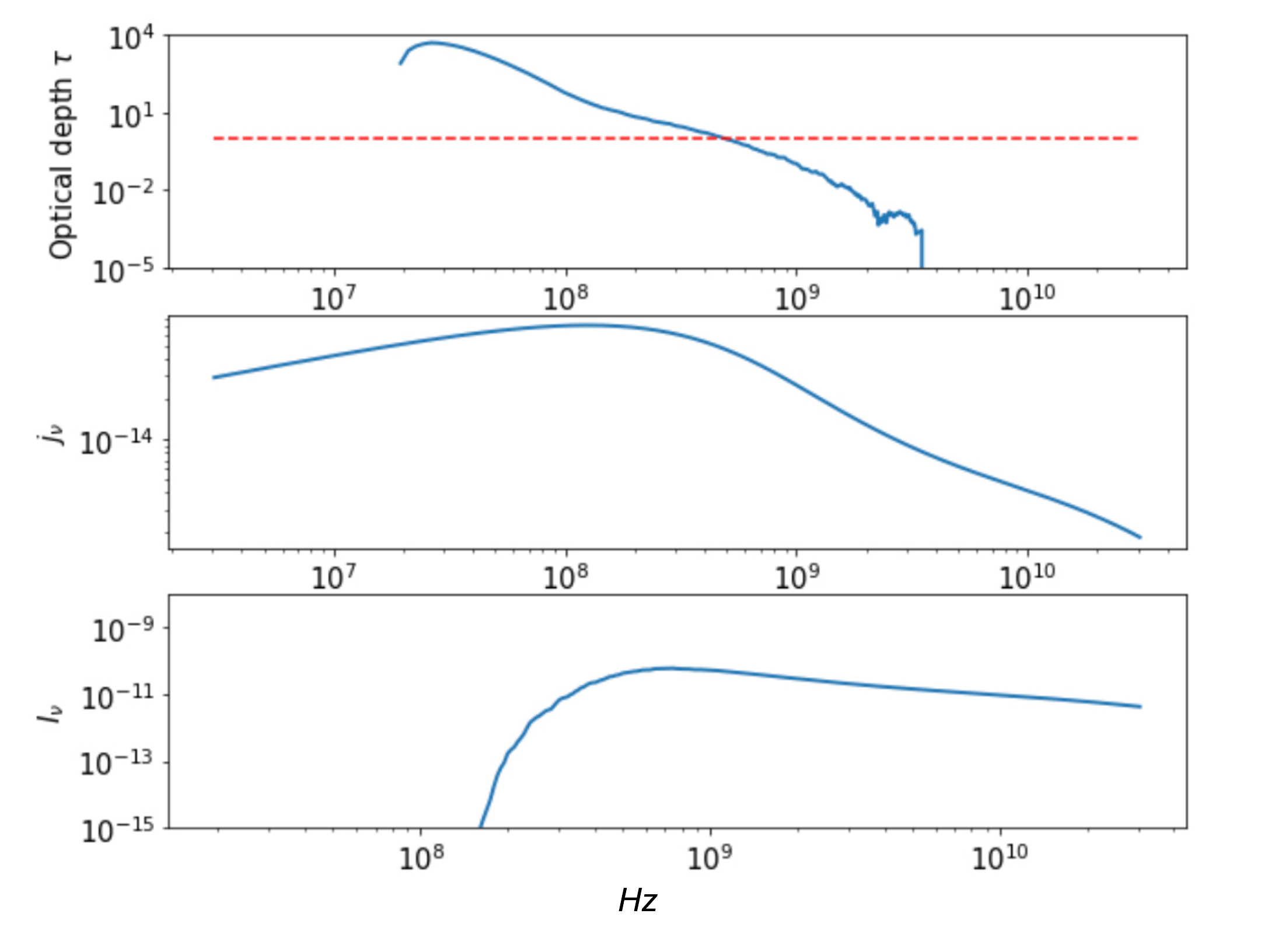}
    \caption{Optical depth $\tau_{\nu,\textrm{SSA}} = d_\text{shock} \alpha_{\nu,\textrm{SSA}} $, synchrotron emissivity $j_\nu$ and $I_\nu$ for the simulation with mass ratio 100 in the downstream left to the shock. The dashed line indicates the $\tau_{\nu,\textrm{SSA}}=1$ limit, where emission becomes optically thin. }
    \label{fig:tau}
\end{figure}

\begin{figure}[htbp]
\includegraphics[width=3in]{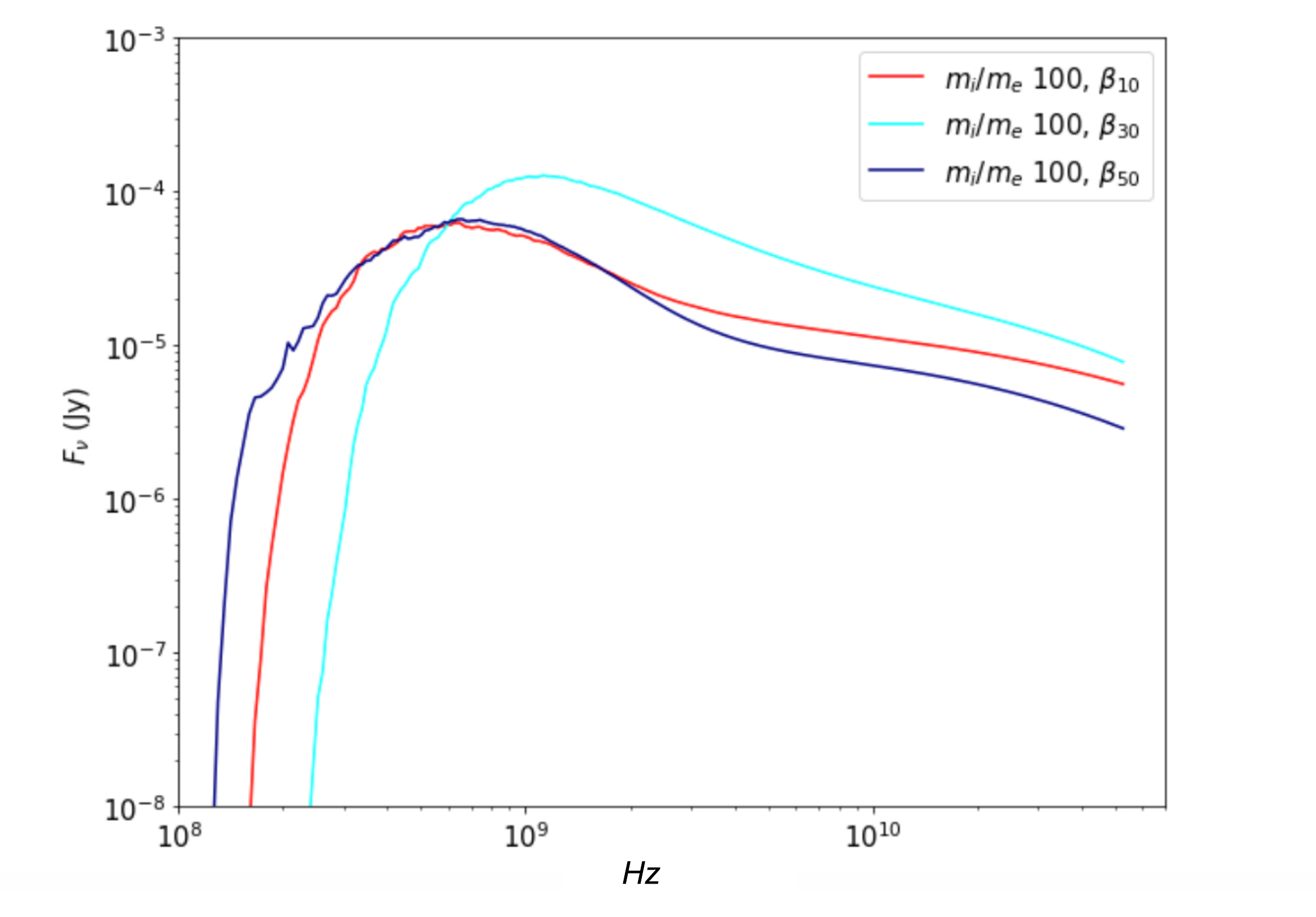}
    \caption{Spectral energy distribution $F_\nu$ for $R_\text{S}=10^{16}$ cm  and a  luminosity distance $l_\text{D} = 1$ Mpc for $\beta$=10,30,50.  }
    \label{fig:specbetas2}
\end{figure}

\begin{figure}[htbp]
\includegraphics[width=3in]{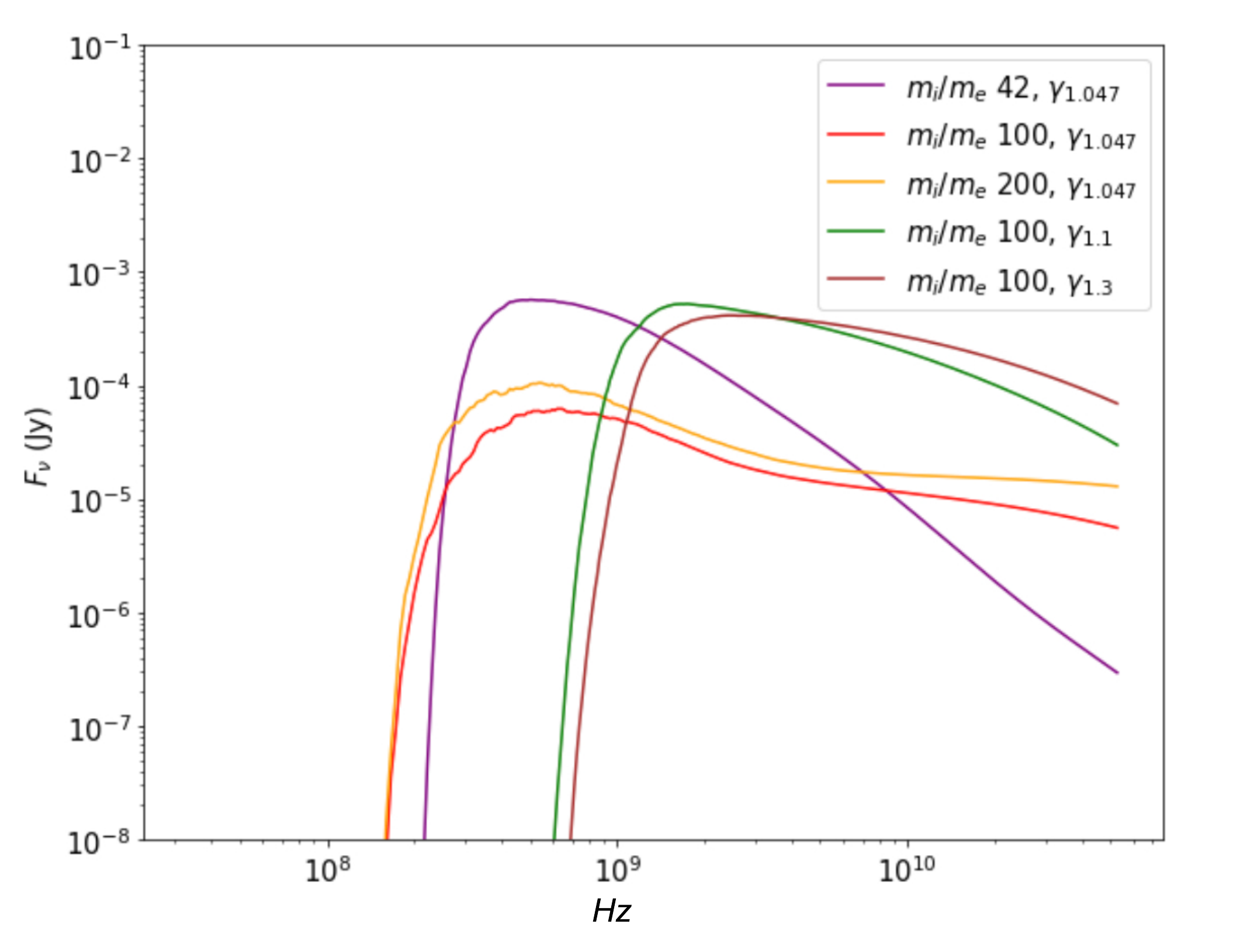}
    \caption{Spectral energy distribution $F_\nu$ for $R_\text{S}=10^{16}$ cm  and a  luminosity distance $l_\text{D} = 1$ Mpc for $\gamma$=1.047,1.1,1.3.  }
    \label{fig:specEnDisgammas}
\end{figure}

Fig. \ref{fig:tau} shows that the source becomes transparent at around $8\times 10^8$ Hz. At this frequency the emission is dominated by non-thermal electrons, while at lower frequencies thermal electrons emit (and absorb) the synchrotron photons. The flux shows a spectral index of about $-1$. Due to the direct correspondence between particle momenta and frequencies the spectrum spans one and a half magnitude as the powerlaw tail of electrons does. Simulations with drastically higher particle numbers could yield a more extended spectrum.

For the arbitrary model parameters (source at 1 Mpc) we reach a flux of multiple 0.1 mJy which is likely to be observed. The peak flux is not inhibited by simulation constraints, i.e., larger simulations or more particle statistics will not change the flux.

Synchrotron cooling has not been taken into account as mentioned above. This will only play a role at much later times and will lead to a break in the spectrum moving to lower frequencies. The cooling time may be estimated, but the position of the break is the product of cooling and (varying) acceleration efficiency. PiC simulations allow for a detailed analysis at a certain time, but no long-time analysis over days and weeks.

\subsection{X-rays from Non-Thermal Emission}

In the previous section radio emission through the synchrotron process has been described. The emission from shock acceleration may reach much higher photon energies \citep{1998ApJ...497L..17S}. The formalism presented here is capable of accounting for this emission, but particle-in-cell simulations are limited in predicting the particle distribution at highest energies. This is on the one hand limited by statistics: In order to cover the particle distribution at energies one order of magnitude higher the particle density would have to increase more than 100 fold. On the other hand the maximum energy of particles is also governed by a larger volume around the shock not covered in the small subvolume ascertainable by particle-in-cell simulations.

From the theory of particle acceleration we may expect to see particles accelerated to higher energies by the mechanism explored here and as consequence photons of higher energies~\citep{2023MNRAS.518.2102S}. However, many of these models rely on low electron spectrum indices (2-2.5).  While appropriate for highly-relativistic jets, as we have shown here, such low powers are not produced in the mildly-relativistic kilonova outflows.  We do, however, have no direct hints from the simulation what would be the highest energy. It is in turn also futile to predict inverse Compton scenarios where synchrotron photons are upscattered by the accelerated electrons \citep{2008ARA&A..46...89R,2022Natur.612..236M}. The simulations at hand are not in contradiction to such a model, but due to the low electron energies in the simulation the calculation of the inverse Compton flux is not feasible.

\section{Conclusions}

The emission from neutron star mergers is powered by a number of sources:  nonthermal emission from the relativistic jet and its ensuing afterglow, thermal emission from the material ejected from the merger and disk wind and the remnant emission as the ejecta propagates through the interstellar medium.  In this paper, we focused on this latter ejecta remnant evolution and emission.   Using equilibrium remnant (Euler) calculations as a guide for the initial conditions, we have run a suite of particle-in-cell (using VPIC) calculations of the shock propagation.  Although these calculations slowly converge toward a $\gamma=5/3$ equilibrium shock, the deviations from the Eulerian shock solution can dramatically alter the emission.

We study the dependence of the shock evolution on a range of shock properties (orientation and energy in the magnetic field, velocity and density of the shock).  We determine the fraction of electrons that are shock-accelerated into a power-law distribution which, depending on the model, ranges from 3-70\%.  The inferred power law ($p$) of this distribution is higher ($-p$ more negative) than the power-law distributions used for gamma-ray burst afterglows.  The canonical power $p$ is roughly 2.2-2.3 for gamma-ray bursts~\citep[although see][]{2006MNRAS.371.1441S}.  The power for our electron distributions are $2.6-4.2$ for a $60^{\circ}$ angle orientation of the magnetic field.  These higher values of the power are expected for our non- or mildly-relativistic outflows of the ejecta remnant.  As the shock becomes non-relativistic, we expect this value to rise to $\sim 4$ \citep{Bell_1978,Blandford_1978, Drury_1983, Blandford_1987} or even softer once we are no longer in the strong shock limit and the compression ratio is appreciably reduced.

We note that the power is very sensitive both physical (magnetic field orientations) and numerical assumptions.  We have focused on the $60^{\circ}$ angular orientation with respect to the shock, but the power-law distribution could be much steeper resulting in fewer high energy electrons with different magnetic field orientations.  An important commonly-used numerical assumption  is the reduction of the proton-to-electron mass ratio.  To make calculations tractable, most simulations reduce this ratio from the actual value of 1836.2.
Somewhat higher mass ratios than used here would be possible for the quasi-onedimensional setup used here, but runs at the natural mass ratio are prohibitively expensive.
If the chosen value is too low (corresponding to a larger effective electron mass), the number of electrons accelerated in the shock to high energies is reduced.  Our high values are, in part, due to the approximations in our VPIC calculations. 

The higher powers in the electron distribution mean that the emission from the ejecta remnant are strongest at longer wavelengths. We find that the X-ray flux from these ejecta remnants is negligible. The radio signal signal would be detectable for a source at a fiducial distance of 1~Mpc at about one day after the kilonova when it is $R_\text{S} = 10^{16}$~cm in radius within seconds with the Green Bank radio telescope. Using \cite{lorimer2005handbook,GBO23b} we computed that the observing time required for a detection at 5~sigma significance is just 28~seconds for a flux of 0.8~mJy at 1.4 GHz. Observing conditions permitting it might even be possible to detect the source with 5 sigma significance in just 7~seconds at a flux of 10~$\mu$Jy in the $38.2--49.8$~GHz band. To obtain a spectral index a somewhat longer observation is required, but 15 to 20~minutes should be sufficient. However for a larger distance of 40Mpc, the flux at 1.4~Ghz would be reduced to 0.5~$\mu$Jy and the time for a detection would jump up to an unmanageable 200~days.

\acknowledgments

The simulations in this work were performed with LANL Institutional Computing which is
supported by the U.S. Department of Energy National
Nuclear Security Administration under Contract No.
89233218CNA000001.
This work was supported in part by the U.S. Department of Energy, Office of Science, Office of Workforce Development for Teachers and Scientists (WDTS) under the Science Undergraduate Laboratory Internship (SULI) program.
MR wants to thank Jonah Miller for helpful discussions and advice on all aspects of scientific computing.
PK wants to thanks Natalia Lewandowska for discussions on the visibility of the predicted synchrotron flux with existing radio telescopes.
The authors would like to thanks Chengkun Huang for advice and discussions on the simulations of collisionless shocks with VPIC.

\appendix

\section{Calculation of the optical thickness}

\label{app:alpha}

We will use Eq.~(6.46) from \citet{rybicki}
\begin{align}
	\alpha_\nu = \frac{c^2}{8\pi h \nu^3} \int \dd^3 p_2 \left( f(p_2^*) - f(p_2) \right) P(\nu,E_2)\,.
	\label{eq:ssa}
\end{align}
Here $p_2$ is the momentum of the electron before emitting a synchrotron photon and $p_2^*$ is the momentum related to the energy $E_2 - h\nu$. This equation is derived under the assumption of an isotropic electron distribution $f(p)$.

Let's examine $p_2^*$ in more detail:
\begin{align}
	E_2^2 &= p_2^2 c^2 + m^2 c^4\\
	(E_2 - h \nu)^2 &= p_2^{*2} c^2 + m^2 c^4\\
	p_2^{*2}  &= \left(\sqrt{p_2^2 c^2 + m^2 c^4} - h\nu\right)^2  - m^2 c^2\\
	\frac{p_2^{*2}}{m^2 c^2} &= \frac{1}{c^2}\left(\sqrt{\frac{p_2^2}{m^2 c^2}+1}-\frac{h\nu}{m c^2}\right)^2 -1\,.\label{eq:p2stern}
\end{align}
The complicated form of this expression can be simplified in the ultra-relativistic case, where $E = p c$. 

We will estimate the difference $f(p_2^*) - f(p_2)$ by a derivative
\begin{align}
	f(p_2^*) - f(p_2) = (p_2^* - p_2)\left.\frac{\partial f}{\partial p}\right\vert_{p=p_2} .
\end{align}
This is obviously only true, when the momentum loss per photon emission is comparably small (this excludes the ultrarelativistic limit). With Eq. \eqref{eq:p2stern} Eq. \eqref{eq:ssa} now reads
\begin{align}
	\alpha_\nu = \frac{c^2}{8\pi h \nu^3} \int \dd p_2\ 4\pi p_2^2\ (p_2^* - p_2)\left.\frac{\partial f}{\partial p}\right\vert_{p=p_2} P(\nu,E_2)\,.
\end{align}
The spectral power density becomes in the delta-approximation (or monochromatic approximation)
\begin{align}
	P_\nu(\gamma) &\approx \beta_s \left(1-\frac{1}{\gamma^2})\right)\delta(\nu-\nu_c)\\
	&= \beta_s \frac{p^2}{m^2 c^2 + p^2}\label{eq:syncemit}\\
	\beta_s &= \frac{2}{3}\frac{e^4 B^2}{m^2 c^3}\\
	\nu_c &= \frac{3}{4}\frac{e B}{\pi m c}\gamma^2\\
	&= \frac{3}{4}\frac{e B}{\pi m c}\left( 1+ \frac{p^2}{m^2 c^2} \right)\,.
\end{align}
Evaluating the delta distribution using the transformation rule for delta distributions returns: 
\begin{align}
	\delta(\nu - \nu_c) &= \left(\left\vert\frac{\partial \nu_c(p)}{\partial p}\right\vert_{p=p_0}\right)^{-1}\delta(p-p_0)\\
	p_0 &= m c\, \sqrt{\frac{4\pi m c}{3 e B}\nu-1}\\
	\frac{\partial \nu_c}{\partial p} &= \frac{3}{2}\frac{e B}{\pi m^3 c^3} p\\
	\frac{\partial \nu_c}{\partial p}\biggr \vert_{p=p_0} &= \frac{3 e B}{2 \pi m^2 c^2}\sqrt{\frac{4\pi m c }{3 e B}\nu-1}\,.\label{eq:delta}
\end{align}
This then yields a lengthy expression for the synchrotron self-absorption coefficient $\alpha_{\nu,\textrm{SSA}}$, which depends on the derivative of the distribution function with respect to the momentum $p$: 
\begin{align}
\alpha_{\nu,\textrm{SSA}} &= \left.\frac{\partial f}{\partial p}\right\vert_{p=p_0} \Theta\left(p_0\right) \frac{c  e^2 (4 \pi   m c \nu -3 e B)^2}{54 h \nu^4}\nonumber \\
& 
\left(-(m c)^2 +m \sqrt{\frac{e B}{4 \pi  m c \nu -3 e B}} \sqrt{\frac{4 \pi m^3  c^5 \nu }{e B}-4 c h \nu  \sqrt{\frac{3 \pi \nu (m c)^3}{e B}}-3 m^2 c^4+3 (h \nu)^2}\right)
\end{align}


\bibliography{refs}{}




\end{document}